\documentclass[
    reprint, aps, prd, amsmath, amssymb, superscriptaddress,
]{revtex4-2}

\usepackage{graphicx}
\usepackage{dcolumn}
\usepackage{bm}
\usepackage[dvipsnames]{xcolor}

\usepackage[normalem]{ulem}

\begin{document}


\title{Conformal-Mapping Method for Horizon Multipoles in Numerical Relativity: \\ Implementation, Kerr Validation, and Applications Beyond Axisymmetry}

\author{Yeong-Bok Bae}
\thanks{
astrobyb@gmail.com, gwkang@cau.ac.kr}
\author{Young-Hwan Hyun}
\email{Corresponding author: younghwan.hyun@gmail.com}
\author{Gungwon Kang}
\thanks{
astrobyb@gmail.com, gwkang@cau.ac.kr}
\affiliation{Department of Physics, Chung-Ang University, Seoul 06974, Republic of Korea}

\date{\today}

\begin{abstract}
We present a numerical method that constructs geometrically defined coordinates on black-hole horizons, from which the multipole moments are computed without assuming axisymmetry. This method, which we denote the conformal-mapping method (CMM), provides a numerical realization of the conformal construction proposed by Ashtekar~\textit{et al.} in 2022, combining discrete Ricci flow, spectral embedding onto the unit sphere, and Möbius gauge fixing by the vanishing-area-dipole condition. We first test the CMM against analytic Kerr benchmarks, and then apply it to an equal-mass, non-spinning binary black-hole merger. We also compare it with an approximate-symmetry-based method. The CMM allows the multipole moments to be expressed in a fixed reference frame, whereas the symmetry-adapted frame can reorient abruptly when the preferred approximate axis changes. In a frame aligned with the orbital angular momentum, the amplitude of the quadrupole mode grows during inspiral and decays after merger, displaying a qualitative ringdown behavior. These results show that the CMM is a useful tool for studying horizon geometry in dynamical situations where no stable symmetry axis is available.
\end{abstract}

\maketitle

\section{Introduction}
\label{sec:introduction}

The detection of gravitational waves from binary black hole mergers~\cite{Abbott:2016} has opened a direct observational window into the strong-field dynamics of general relativity. The observed waveforms carry imprints of the highly curved spacetime geometry near the merging black holes, and extracting physical information from these imprints requires a detailed understanding of the near-horizon dynamics in the strong-field regime. A natural way to access this information is through the intrinsic geometry of the black-hole horizon, which provides a quasi-local characterization of the strong-field region near the merging black holes. Multipole moments on horizon cross-sections compactly characterize this geometry, with mass multipoles capturing the intrinsic 2-metric of the cross-section and angular momentum multipoles capturing the rotational structure carried by the horizon's null generators. Recent studies have demonstrated that horizon deformations are closely correlated with gravitational radiation~\cite{Prasad:2024,Bae:2024,RibesMetidieri:2025}, reinforcing the motivation for developing robust methods to compute these moments in numerical simulations. In particular, close black-hole scattering provides a complementary strong-field setting in which ringdown radiation can already encode detailed information about transient horizon distortions, thereby emphasizing the need for robust quasi-local diagnostics of horizon dynamics~\cite{Bae:2024}.

A systematic definition of horizon multipole moments was given by Ashtekar, Engle, Pawlowski, and Van Den Broeck~\cite{Ashtekar:2004} for axisymmetric isolated horizons, and related dynamical-horizon tools were implemented in numerical relativity to study time-dependent black holes and their horizon multipoles~\cite{Schnetter:2006}. In this framework, axial symmetry provides a preferred axis and associated coordinate system on the horizon, enabling a spherical harmonic decomposition of the Weyl scalar $\Psi_{2}$ that yields two families of gauge-invariant multipole moments: the shape multipole moments $I_\ell$, encoding the intrinsic scalar curvature, and the current multipole moments $L_\ell$, encoding the rotational structure.  This axisymmetric definition has been widely applied in numerical relativity to study the dynamics of individual horizons during inspiral~\cite{Prasad:2022,Prasad:2024} and the approach to equilibrium of the remnant horizon~\cite{Ashtekar:2013,Gupta:2018,Chen:2022}. Ashtekar, Campiglia, and Shah~\cite{Ashtekar:2013} formulated a dynamical-horizon multipole framework for studying the approach to the final Kerr state, in which the late-time axisymmetric structure provides a reference basis for tracking horizon multipoles. In particular, Chen~\textit{et al.}~\cite{Chen:2022} constructed spatially gauge-invariant multipole moments on the common horizon by Lie-dragging the spherical-harmonic basis from the late-time axisymmetric state, and showed that the resulting moments are well described by quasinormal modes.

In practice, computing multipole moments on numerical horizons requires constructing a preferred coordinate system on each horizon surface. Several approaches have been developed within the axisymmetric framework. The Killing transport method~\cite{Dreyer:2003,Ashtekar:2004} uses the Killing transport equations to construct a preferred axial vector field on the horizon. The approximate Killing vector (AKV) method~\cite{Cook:2007,Owen:2009} instead adopts a variational approach, identifying the vector field that best approximates a rotational symmetry by minimizing a strain functional over divergence-free fields on the 2-surface. Both methods rely on the existence of an approximate axial symmetry and therefore become ambiguous or unstable when the horizon geometry departs significantly from axisymmetry. This can occur even during inspiral when different deformations are associated with different directions; for example, spin-induced and tidal distortions are naturally associated with the spin and companion directions, respectively. The ambiguity becomes more pronounced near merger as tidal distortions strengthen, and also in strongly distorted non-merger encounters such as close black-hole scattering~\cite{Bae:2024}.

More recently, Ashtekar, Khera, Kolanowski, and Lewandowski~\cite{Ashtekar:2022} proposed a method to define horizon multipole moments for generic non-expanding horizons that does not require any symmetry assumption. Building on the earlier work of Korzy\'{n}ski~\cite{Korzynski:2007}, who used the conformal decomposition of the horizon metric to define quasi-local angular momentum without assuming axisymmetry, this framework constructs a unit round metric in the conformal class of the physical horizon metric. The three-parameter conformal freedom is fixed by requiring the area dipole moment to vanish, yielding a unique conformal round metric and an associated spherical-harmonic basis up to an overall rotation, and hence a set of multipole moments $I_{\ell m}$ and $L_{\ell m}$ defined up to the corresponding rotation within each $\ell$ sector. Gourgoulhon~\textit{et al.}~\cite{Gourgoulhon:2026} recently applied both definitions to the Kerr horizon, deriving closed-form results for the axisymmetric moments and key conformal quantities, and comparing the two resulting multipole families in detail.

Despite its theoretical appeal, the conformal construction of Ref.~\cite{Ashtekar:2022} has not previously been implemented numerically. The purpose of this paper is to fill this gap.  We present a complete numerical pipeline for computing horizon multipole moments based on this conformal framework, which we refer to as the conformal-mapping method (CMM). Our approach consists of determining the conformal factor via a discrete Ricci-flow--type procedure, constructing a spherical-harmonic basis on the resulting unit round metric, and fixing the residual M\"{o}bius freedom by imposing the vanishing of the area dipole moment.  We validate the implementation against Kerr analytic benchmarks, including the corrected conformal map and the associated multipoles, and apply it to an equal-mass, non-spinning binary black hole merger simulation, comparing the results with those obtained from the symmetry-based method.

The paper is organized as follows. Section~\ref{sec:theory} reviews the theoretical framework for the
axisymmetric and conformal horizon multipole constructions, their relation in the
axisymmetric limit, and the Kerr benchmarks used for validation. Section~\ref{sec:implementation} describes the numerical implementation of the CMM and introduces the AKV method as a
symmetry-based prescription used for comparison. Section~\ref{sec:kerr} presents the validation against analytic Kerr solutions.  Section~\ref{sec:bbh} applies both methods to a binary black hole merger and compares the results.  Section~\ref{sec:discussion} summarizes our main results and discusses future directions.

\section{Conformal methods for horizon multipoles}

\subsection{Theoretical Framework}
\label{sec:theory}

The purpose of horizon multipoles is to provide a quasi-local and coordinate-independent
description of the geometry of a black-hole horizon cross-section~\cite{Ashtekar:2004,Ashtekar:2022}. As in Ref.~\cite{Ashtekar:2022}, the relevant geometric object is a smooth spacelike 2-surface $\mathcal S \simeq S^2$ on the non-expanding horizon $\Delta\simeq S^2\times \mathbb{R}$. Its intrinsic geometry is encoded in the physical
2-metric $q_{ab}$, the associated area element $d^2V$, the scalar curvature
$\mathcal R$, and the pullback $\omega_a$ of the horizon rotation 1-form.
Let
\begin{equation}
  A_\Delta := \oint_{\mathcal S} d^2V,
  \qquad
  R_\Delta := \sqrt{\frac{A_\Delta}{4\pi}} ,
  \label{eq:theory_areal_radius}
\end{equation}
where $A_\Delta$ is the area of the horizon cross-section $\mathcal S$ and $R_\Delta$ its areal radius.
For vacuum non-expanding and isolated horizons, the intrinsic curvature
and rotational data are related to the Newman--Penrose Weyl scalar $\Psi_2$ by
\begin{equation}
   -\frac{\mathcal R}{4}=\mathrm{Re}\,\Psi_2,
  \qquad
  D_{[a}\omega_{b]} = \mathrm{Im}\,\Psi_2\,\epsilon_{ab},
  \label{eq:theory_Psi2_relation}
\end{equation}
where $D_a$ is the Levi-Civita connection of $q_{ab}$ and
$\epsilon_{ab}$ the area 2-form. It is therefore natural to combine the mass-type
and current-type horizon data into the complex seed
\begin{equation}
  \Phi_\Delta := \frac{{\mathcal R}}{4} - i\,\mathrm{Im}\,\Psi_2 .
  \label{eq:theory_seed}
\end{equation}
In vacuum, $\Phi_\Delta = -\Psi_2$. The multipole problem is then to expand
$\Phi_\Delta$ in a geometrically preferred spherical-harmonic basis on $\mathcal S$. 

A systematic construction of horizon multipoles for axisymmetric isolated horizons was
given in Ref.~\cite{Ashtekar:2004}. If $\mathcal S$ admits an axial
Killing field $\varphi^a$ with closed orbits and two fixed points, the symmetry determines
a canonical latitude function $\zeta \in [-1,1]$ through
\begin{equation}
  D_a \zeta
  =
  \frac{1}{R_\Delta^2}\,\epsilon_{ba}\varphi^b,
  \qquad
  \oint_{\mathcal S}\zeta\, d^2V = 0 .
  \label{eq:theory_zeta_def}
\end{equation}
Together with an affine angle $\varphi$ along the Killing orbits, this yields the canonical
axisymmetric form of the physical metric
\begin{align}
q_{ab}dx^a dx^b
 &=
 R_\Delta^2
 \left[
    f(\zeta)^{-1} d\zeta^2 + f(\zeta)\, d\varphi^2
 \right],
 \nonumber\\
d^2V &= R_\Delta^2\, d\zeta\, d\varphi .
 \label{eq:theory_axisym_metric}
\end{align}
Using the same invariant coordinates $(\zeta,\varphi)$, one then introduces the auxiliary
canonical round metric by replacing $f(\zeta)$ with the round-sphere function
$1-\zeta^2$:
\begin{equation}
  \tilde{\mathring q}_{ab}dx^a dx^b
  =
  R_\Delta^2
  \left[
    \frac{d\zeta^2}{1-\zeta^2} + (1-\zeta^2)\, d\varphi^2
  \right].
  \label{eq:theory_axisym_round_metric}
\end{equation}
This metric is round and has the same area element as the physical metric. Its associated
$m=0$ spherical harmonics are
\begin{equation}
  \tilde{\mathring Y}_{n0}(\zeta)
  =
  \sqrt{\frac{2n+1}{4\pi}}\,P_n(\zeta),
  \label{eq:theory_axisym_Yn0}
\end{equation}
with $P_n$ the Legendre polynomials. Axisymmetric geometric multipoles are then
\begin{equation}
  I_n^{(\mathrm{axi})} + iL_n^{(\mathrm{axi})}
  :=
  \oint_{\mathcal S}
  \Phi_\Delta\,\tilde{\mathring Y}_{n0}\, d^2V ,
  \label{eq:theory_axi_geometric}
\end{equation}
or, equivalently,
\begin{align}
 & I_n^{(\mathrm{axi})}
  =
  \oint_{\mathcal S}\frac{{\mathcal R}}{4}\,\tilde{\mathring Y}_{n0}\, d^2V,
\nonumber\\
& L_n^{(\mathrm{axi})}
  =
  -\oint_{\mathcal S}\mathrm{Im}\,\Psi_2\,\tilde{\mathring Y}_{n0}\, d^2V .
  \label{eq:theory_axi_I_L}
\end{align}
These are dimensionless geometric moments. The corresponding dimensionful mass and
angular-momentum multipoles are
\begin{align}
&M_n^{(\mathrm{axi})}
  =
  \sqrt{\frac{4\pi}{2n+1}}\,
  \frac{M_\Delta R_\Delta^n}{2\pi}\, I_n^{(\mathrm{axi})},
  \nonumber\\
&J_n^{(\mathrm{axi})}
  =
  \sqrt{\frac{4\pi}{2n+1}}\,
  \frac{R_\Delta^{n+1}}{4\pi G}\, L_n^{(\mathrm{axi})}.
  \label{eq:theory_axi_Mn_Jn}
\end{align}
With this normalization,
\begin{equation}
M_0^{(\mathrm{axi})}=M_\Delta,
\qquad
J_1^{(\mathrm{axi})}=J_\Delta,
\label{eq:theory_axi_anchors}
\end{equation}
where $M_\Delta$ and $J_\Delta$ denote the horizon mass and angular momentum, respectively.
The corresponding geometric moments satisfy
\begin{equation}
I_0^{(\mathrm{axi})}=\sqrt{\pi},
\qquad
L_0^{(\mathrm{axi})}=0,
\qquad
I_1^{(\mathrm{axi})}=0,
\label{eq:theory_axi_constraints}
\end{equation}
so the axisymmetric construction is automatically mass-centered.

The central limitation of the axisymmetric construction is equally clear: the preferred
latitude $\zeta$ is defined from the axial Killing field itself. Once the horizon is no longer
axisymmetric, there is no preferred rotational vector, no preferred latitude, and hence no
canonical harmonic basis of this type. This is precisely the situation encountered in generic
binary black-hole spacetimes, especially near merger, when the common horizon can become
strongly distorted and no exact or even reliable approximate axisymmetry need be present.
A more general construction is therefore needed if one wants multipoles that depend only on
the intrinsic horizon geometry.

Such a construction was developed for generic
non-expanding horizons~\cite{Ashtekar:2022}. The key idea is to replace the symmetry-adapted auxiliary round
metric $\tilde{\mathring q}_{ab}$ by a canonical unit round metric $\mathring q_{ab}$ obtained
from the conformal class of the physical metric. Since every smooth metric on $S^2$ is
conformal to a round one, one may write
\begin{equation}
  \mathring q_{ab} = \psi^2  q_{ab},
  \label{eq:theory_conformal_round}
\end{equation}
where $\mathring q_{ab}$ is a unit round metric. In two dimensions, the scalar-curvature
transformation law gives the nonlinear elliptic equation
\begin{equation}
  D^2 \ln\psi + \psi^2 = \frac{1}{2}{\mathcal R}.
  \label{eq:theory_psi_equation}
\end{equation}
Thus the problem of constructing a preferred spherical-harmonic basis is first reduced to the
problem of selecting a preferred round metric in the conformal class of $q_{ab}$.

However, the conformal round metric is not unique. There remains the three-parameter
M\"obius, or Lorentz-boost, freedom among unit round metrics in the same conformal class:
\begin{align}
  \mathring q'_{ab} &= \alpha^2 \mathring q_{ab},
\quad
  \alpha^{-1}
  = \alpha_0 + \sum_{i=1}^{3}\alpha_i r^i,
  \quad
  \alpha_0^2 - |\vec\alpha|^2 = 1 ,
  \label{eq:theory_alpha_family}
\end{align}
where $r^i$ are the standard $\ell=1$ embedding functions on the round sphere. The
conformal construction fixes this freedom by requiring the area dipole to vanish:
\begin{equation}
  \oint_{\mathcal S}\mathring Y_{1m}\, d^2V = 0,
  \qquad
  m=-1,0,1,
  \label{eq:theory_area_dipole_zero}
\end{equation}
where $\mathring Y_{\ell m}$ are the spherical harmonics of the chosen unit round metric
$\mathring q_{ab}$. This selects a unique conformal round frame for any smooth horizon metric, modulo an
overall $\mathrm{SO}(3)$ rotation of the resulting spherical-harmonic basis.
Geometrically, the condition plays the role of a center-of-mass condition for the positive area
measure on the horizon.

The positivity of the area measure is important. The uniqueness argument relies on it, and
this is precisely why the vanishing-area-dipole condition is technically robust. By contrast,
a curvature-weighted mass-dipole condition would not have the same status on a generic
distorted horizon because the curvature weight need not be positive definite. Thus the
conformal construction does not, in general, force the mass dipole to vanish; rather, it selects
the unique round frame in which the \emph{area} dipole vanishes. This is one of the main
conceptual differences between the axisymmetric and conformal constructions.

Once the canonical unit round metric $\mathring q_{ab}$ has been fixed, its spherical
harmonics $\mathring Y_{\ell m}$ provide the preferred basis on the physical horizon. The
generic geometric multipoles are then defined by
\begin{equation}
  I_{\ell m}^{(\mathrm{conf})} + iL_{\ell m}^{(\mathrm{conf})}
  :=
  \oint_{\mathcal S}
  \Phi_\Delta\, \mathring Y_{\ell m}\, d^2V ,
  \label{eq:theory_conf_geometric}
\end{equation}
that is,
\begin{align}
&I_{\ell m}^{(\mathrm{conf})}
 :=
 \oint_{\mathcal S}\frac{{\mathcal R}}{4}\,\mathring Y_{\ell m}\, d^2V,
 \nonumber\\
&L_{\ell m}^{(\mathrm{conf})}
 :=-
 \oint_{\mathcal S}\mathrm{Im}\,\Psi_2\,\mathring Y_{\ell m}\, d^2V .
 \label{eq:theory_conf_I_L}
\end{align}
The corresponding dimensionful multipoles are
\begin{align}
&M_{\ell m}^{(\mathrm{conf})}
 =
 \sqrt{\frac{4\pi}{2\ell+1}}\,
 \frac{M_\Delta R_\Delta^\ell}{2\pi}\, I_{\ell m}^{(\mathrm{conf})},
 \nonumber\\
&J_{\ell m}^{(\mathrm{conf})}
 =
 \sqrt{\frac{4\pi}{2\ell+1}}\,
 \frac{R_\Delta^{\ell+1}}{4\pi G}\, L_{\ell m}^{(\mathrm{conf})}.
 \label{eq:theory_conf_Mlm_Jlm}
\end{align}
In the present work, we focus mainly on the mass-type sector and, in practice, will mostly use
the geometric moments $I_{\ell m}^{(\mathrm{conf})}$.

A useful point is that the monopole is fixed by topology. Since $\mathcal S \simeq S^2$,
Gauss--Bonnet gives
\begin{equation}
  \oint_{\mathcal S} {\mathcal R}\, d^2V = 8\pi.
  \label{eq:theory_gauss_bonnet}
\end{equation}
Thus the dimensionless shape monopole is fixed topologically, $I_{00}=\sqrt{\pi}$, and carries no information about the horizon distortion; the nontrivial shape information begins in the higher multipoles.

It is important to emphasize that the conformal construction does not simply reduce
to the axisymmetric one when the horizon happens to be axisymmetric. If the physical metric $q_{ab}$ admits an axial Killing field, then the conformal unit
round metric $\mathring q_{ab}$ selected by Eq.~\eqref{eq:theory_area_dipole_zero} is
invariant under the same axial Killing flow on $\mathcal S$~\cite{Ashtekar:2022}. Equivalently, the two metrics
share the same family of closed rotational Killing orbits. What changes is not the orbit
structure itself, but the preferred latitude used to parametrize the orbit space. In the 
axisymmetric construction this latitude is the area-fraction coordinate $\zeta$, whereas in
the conformal construction it is the polar coordinate associated with the conformal
unit round metric. Therefore the two multipole families need not coincide even when they are
built from the same axisymmetric horizon geometry.

To make this relation explicit, choose coordinates $(Z,\varphi)$ adapted to the common
axial Killing flow such that the canonical unit round metric takes the form
\begin{equation}
  \mathring q_{ab}dx^a dx^b
  =
  \frac{dZ^2}{1-Z^2} + (1-Z^2)\, d\varphi^2,
  \quad
  Z=\cos\vartheta \in [-1,1].
  \label{eq:theory_round_metric_Z}
\end{equation}
Then the physical metric may be written as
\begin{align}
 q_{ab}dx^a dx^b
  &=
  \psi^{-2}(Z)
  \left[
    \frac{dZ^2}{1-Z^2} + (1-Z^2)\, d\varphi^2
  \right],
 \nonumber\\
  d^2V &= \psi^{-2}(Z)\, dZ\, d\varphi .
  \label{eq:theory_physical_metric_Z}
\end{align}
From the same physical metric, one reconstructs the axisymmetric latitude using the area-fraction relation $\frac{d\zeta}{dZ} = \frac{\psi^{-2}(Z)}{R_\Delta^2}$:
\begin{equation}
  \zeta(Z)
  :=
  -1 + \frac{1}{R_\Delta^2}\int_{-1}^{Z}\psi^{-2}(\bar Z)\, d\bar Z.
  \label{eq:theory_zeta_of_Z}
\end{equation}
Substituting Eq.~\eqref{eq:theory_zeta_of_Z} into
Eq.~\eqref{eq:theory_physical_metric_Z} reproduces the canonical axisymmetric form
\begin{align}
  q_{ab}dx^a dx^b
  &=
  R_\Delta^2
  \left[
    f(\zeta)^{-1} d\zeta^2 + f(\zeta)\, d\varphi^2
  \right],
  \nonumber\\
 f\bigl(\zeta(Z)\bigr)
  &=
  \frac{(1-Z^2)\psi^{-2}(Z)}{R_\Delta^2}.
  \label{eq:theory_f_from_Z}
\end{align}
Thus the two descriptions are related by an explicit monotone map
$Z \leftrightarrow \zeta$.

This is the precise sense in which the axisymmetric specialization of the conformal
construction is related to the axisymmetric framework. Both are one-dimensional
projections of the same physical seed $\Phi_\Delta$ with the same physical measure; the
difference lies entirely in the harmonic weight:
\begin{equation}
  I_n^{(\mathrm{axi})} + iL_n^{(\mathrm{axi})}
  =
  2\pi R_\Delta^2
  \int_{-1}^{1}
  \Phi_\Delta(\zeta)\,
  \tilde{\mathring Y}_{n0}(\zeta)\,
  d\zeta ,
  \label{eq:theory_axi_1d}
\end{equation}
whereas
\begin{equation}
  I_{n0}^{(\mathrm{conf})} + iL_{n0}^{(\mathrm{conf})}
  =
  2\pi R_\Delta^2
  \int_{-1}^{1}
  \Phi_\Delta(\zeta)\,
  \mathring Y_{n0}\!\bigl(Z(\zeta)\bigr)\,
  d\zeta .
  \label{eq:theory_conf_axisym_1d}
\end{equation}
Since
\begin{equation}
  \tilde{\mathring Y}_{n0}(\zeta)
  =
  \sqrt{\frac{2n+1}{4\pi}}\,P_n(\zeta),
  \quad
  \mathring Y_{n0}(Z)
  =
  \sqrt{\frac{2n+1}{4\pi}}\,P_n(Z),
  \label{eq:theory_Pn_weights}
\end{equation}
the difference between the two constructions reduces to the replacement
\begin{equation}
  P_n(\zeta)\quad\longrightarrow\quad P_n\!\bigl(Z(\zeta)\bigr).
  \label{eq:theory_Pn_replacement}
\end{equation}
Thus, even for an axisymmetric horizon, the multipoles defined by the conformal construction do not in general coincide with those defined by the axisymmetric construction. Although both are built from the same horizon geometry, they use different canonical identifications of the orbit space with the interval $[-1,1]$.

At the same time, the conformal construction may be viewed as providing a more general
starting point from which an axisymmetric-style description can be recovered whenever a
preferred family of closed curves on the horizon is specified. The conformal round
coordinates naturally supply such a family even when the horizon is not exactly
axisymmetric, and one may then define an area-fraction latitude along those curves in direct
analogy with the axisymmetric construction. In a genuinely non-axisymmetric situation, the
resulting axisymmetric-style reduction should be regarded as an auxiliary diagnostic rather
than as an invariant symmetry-based definition. However, when exact axisymmetry is
restored, this auxiliary area-fraction reduction reduces to the usual area-fraction coordinate $\zeta$, and the
corresponding multipoles recover the older axisymmetric moments. In this sense, the
conformal construction not only extends beyond axisymmetry, but also provides a natural
framework in which the symmetry-adapted construction reappears as a special case.

The Kerr horizon provides the cleanest analytic benchmark because both constructions apply.
Let $r_+$ be the outer horizon radius and define
\begin{equation}
  u := \cos\theta_{\rm BL},
  \qquad
  R_\Delta^2 = r_+^2 + a^2 .
  \label{eq:theory_kerr_u}
\end{equation}
On a Boyer--Lindquist horizon cross-section, the physical area element is
$d^2V = R_\Delta^2\, du\, d\varphi$, so the axisymmetric area-fraction coordinate is simply
\begin{equation}
  \zeta = u .
  \label{eq:theory_kerr_zeta_equals_u}
\end{equation}
In the Kinnersley tetrad~\cite{Kinnersley:1969zza},
\begin{equation}
  \mathrm{Re}\,\Psi_2(u)
  =
  -M\,
  \frac{r_+\,(r_+^2 - 3a^2u^2)}{(r_+^2 + a^2u^2)^3}.
  \label{eq:theory_kerr_RePsi2}
\end{equation}
Accordingly, axisymmetric Kerr mass multipoles are
\begin{equation}
  M_n^{(\mathrm{axi})}
  =
  -M_\Delta R_\Delta^{n+2}
  \int_{-1}^{1}
  P_n(u)\,\mathrm{Re}\,\Psi_2(u)\, du,
  \label{eq:theory_kerr_Mn_axi}
\end{equation}
whereas conformal Kerr multipoles are
\begin{equation}
  M_n^{(\mathrm{conf})}
  =
  -M_\Delta R_\Delta^{n+2}
  \int_{-1}^{1}
  P_n\!\bigl(Z(u)\bigr)\,\mathrm{Re}\,\Psi_2(u)\, du .
  \label{eq:theory_kerr_Mn_conf}
\end{equation}
Thus the difference again resides entirely in the replacement of the harmonic weight
$P_n(u)$ by $P_n(Z(u))$.

For Kerr, the corrected conformal-round latitude is
\begin{equation}
  Z(u)
  =
  \tanh\!\left(
    \operatorname{artanh}u - \frac{a^2}{R_\Delta^2}\,u
  \right),
  \label{eq:theory_kerr_Z_of_u}
\end{equation}
and the corresponding conformal factor can be written as
\begin{equation}
  \psi(u)
  =
  \frac{\sqrt{(r_+^2+a^2u^2)\,\bigl(1-Z(u)^2\bigr)}}
       {R_\Delta^2\sqrt{1-u^2}} .
  \label{eq:theory_kerr_psi}
\end{equation}
The Kerr expressions above, including the corrected conformal map relative to Ref.~\cite{Ashtekar:2022}, were part of the analytic groundwork for this implementation. A detailed analytic treatment of the same Kerr construction was recently presented in Ref.~\cite{Gourgoulhon:2026}; here we use these results as benchmarks for the numerical pipeline.
This point is important for the present paper: the Kerr test is not merely a bookkeeping check of spherical harmonics. A correct implementation must recover the conformal map to the conformal round sphere and the corresponding conformal multipoles, while also allowing comparison with the older symmetry-adapted construction in the axisymmetric case. Because $Z(u)\neq u$ for $a\neq0$, the two Kerr multipole families
are genuinely different. For example, in the small-spin limit,
\begin{align}
  &M_2^{(\mathrm{axi})} = -\frac{4}{5}\,Ma^2 + \mathcal{O}(a^4),\nonumber\\
  &M_2^{(\mathrm{conf})} = -\frac{6}{5}\,Ma^2 + \mathcal{O}(a^4).  \label{eq:theory_kerr_quadrupole_compare}
\end{align}
More generally, in the mass sector, equatorial symmetry makes all odd-$n$
moments vanish for Kerr in both constructions. The two definitions therefore
first differ in the quadrupole, $n=2$, for nonzero spin.

The conformal construction also differs in an essential way from symmetry-based numerical
methods such as the Killing transport method or the approximate Killing vector approach~\cite{Dreyer:2003,Ashtekar:2004,Cook:2007,Owen:2009}.
Those methods seek a preferred rotational vector field and then construct an axis-adapted
basis from it. In practice, they can be applied even when the horizon is not exactly
axisymmetric, and have been used successfully in regimes where an approximate rotational
symmetry is present. However, their output is then tied to the choice of approximate
symmetry, and becomes increasingly ambiguous as the horizon departs more strongly from
axisymmetry. By contrast, the conformal construction starts directly from the intrinsic metric
and continues to provide a canonical harmonic basis without requiring a preferred symmetry
axis as an input. That feature is central for the present work: our primary goal is not to
revisit the already-analyzed Kerr geometry, but to demonstrate through an explicit numerical
implementation that the conformal construction provides robust horizon multipoles without
assuming axisymmetry.

In summary, the axisymmetric and conformal constructions expand the same physical seed
$\Phi_\Delta$ on two different canonical harmonic bases. In the axisymmetric case the basis
is determined by the intrinsic latitude $\zeta$ and the auxiliary round metric
$\tilde{\mathring q}_{ab}$ built from the axial symmetry. In the conformal case the basis is
determined instead by the canonical unit round metric $\mathring q_{ab}$ selected by the
vanishing-area-dipole condition. In axisymmetry the relation between the two is encoded
entirely in the map $\zeta \leftrightarrow Z$, once the two descriptions are compared with the same symmetry axis. For the purposes of the present work, this distinction is crucial: only the conformal
construction continues to provide a canonical harmonic basis when no preferred symmetry
axis is available. The conformal construction is therefore the
natural framework for the present paper, whose main purpose is to show numerically that
horizon multipoles can be computed reliably even when the horizon geometry is not
axisymmetric.

\subsection{Numerical Implementation}
\label{sec:implementation}

We now turn to the numerical realization of the conformal construction introduced in Sec.~\ref{sec:theory}. The conformal-mapping method (CMM) builds on the conformal framework of Ref.~\cite{Ashtekar:2022} and develops it into an explicit computational pipeline for horizon cross-sections in numerical relativity. Although the original multipole framework was formulated for non-expanding horizons, the CMM coordinate construction uses only the intrinsic geometry of a smooth spherical cross-section. In the dynamical applications of Sec.~\ref{sec:bbh}, we therefore apply it to marginally outer trapped surfaces (MOTSs) as a quasi-local intrinsic-geometric diagnostic, without assuming that their world tubes are non-expanding or isolated. The input data consist of horizon cross-sections extracted from numerical relativity simulations, which provide the intrinsic 2-metric $q_{AB}$ and the Ricci scalar $\mathcal{R}$ on each cross-section $\mathcal{S}$, defined on a structured grid that we denote $(\theta_{\rm sim}, \phi_{\rm sim})$. In this work we focus on the mass multipole moments, which are determined by the scalar curvature, $\mathcal{R}$.

At the numerical level, the construction has two main parts: (i) building a unit round metric $\mathring q_{ab} = \psi^2 q_{ab}$ conformally related to the physical horizon metric, by solving the discrete analog of Eq.~\eqref{eq:theory_psi_equation}; and (ii) embedding $\mathcal{S}$ onto the unit round 2-sphere, on which the harmonic decomposition is carried out. We denote the resulting numerical coordinates by $(\theta_{\rm CMM}, \phi_{\rm CMM})$; in the continuum limit they reduce to the round-sphere coordinates $(\vartheta, \varphi)$ adapted to $\mathring q_{ab}$ in the theoretical construction of Sec.~\ref{sec:theory}. The spherical-harmonic basis $\mathring Y_{\ell m}$ entering the multipole decomposition is evaluated on $(\theta_{\rm CMM}, \phi_{\rm CMM})$.

For Part~(i), we solve the equivalent problem using discrete Ricci flow on a triangular mesh~\cite{Luo:2004, Springborn:2008}. The underlying geometric idea is simple: starting from the physical horizon metric on $\mathcal{S}$, we deform it by a point-dependent overall rescaling---a conformal factor---that progressively smooths out non-uniformities in the Gaussian curvature, until the surface becomes a unit round sphere. The fixed point of this flow is, by construction, the solution of Eq.~\eqref{eq:theory_psi_equation}. A practical advantage of this formulation is that the flow can be expressed entirely in terms of intrinsic geometric quantities---edge lengths and the angles they subtend---avoiding the coordinate singularities at the poles that arise when Eq.~\eqref{eq:theory_psi_equation} is discretized in $(\theta_{\rm sim}, \phi_{\rm sim})$ directly.

The horizon metric data, originally defined on $(\theta_{\rm sim}, \phi_{\rm sim})$, are interpolated onto a triangulation of the sphere, and the intrinsic physical edge lengths $L_{ij}^{\rm phys}$ are computed from the physical 2-metric. We rescale these to unit areal radius, $L_{ij} := L_{ij}^{\rm phys}/R_{\Delta}$, so that the input mesh has total area approximately $4\pi$. The discrete Ricci flow then seeks per-vertex log conformal factors $\lambda_i$ such that the rescaled edge lengths
\begin{equation}
  \tilde L_{ij} \;=\; L_{ij}\,
    \exp\!\left[\frac{\lambda_i + \lambda_j}{2}\right]
  \label{eq:rescaled-edge-length}
\end{equation}
define a discrete unit round metric. In the discrete setting the Gaussian curvature is concentrated at the vertices: each vertex carries an angle deficit---the amount by which the triangle angles incident to it fail to sum to $2\pi$---and Gauss--Bonnet guarantees that these deficits sum to $4\pi$ over the entire mesh. The discrete unit-round condition is the natural counterpart of constant Gaussian curvature: the deficits are distributed across vertices in proportion to the surface area each vertex represents, and the flow adjusts $\lambda_i$ until this distribution is achieved. In the continuum limit, $\lambda$ corresponds to $\ln(R_{\Delta}\,\psi)$, with $\psi$ the conformal factor of Eq.~\eqref{eq:theory_conformal_round}.

For Part~(ii), the conformal mesh obtained from the Ricci flow is embedded onto the unit round 2-sphere by a spectral method: the three eigenvectors of the discrete Laplacian corresponding to the lowest nontrivial eigenspace approximate the Cartesian coordinate functions on the round sphere, providing an embedding that is robust even near the poles. The residual gauge freedom is then fixed in two steps. First, the three-parameter M\"obius boost freedom within the conformal class is removed by imposing the vanishing-area-dipole condition of Eq.~\eqref{eq:theory_area_dipole_zero}, as prescribed in Ref.~\cite{Ashtekar:2022}. Second, the residual $\mathrm{SO}(3)$ rotational freedom is fixed by an external convention appropriate to the physical problem. For the binary application considered below, we align the polar axis with the orbital angular momentum and set the origin of the azimuth by projecting a fiducial simulation axis onto the equatorial plane. This step fixes only the overall orientation of the round-sphere coordinates, completing the construction of $(\theta_{\rm CMM}, \phi_{\rm CMM})$.

For comparison, we also implement the AKV method~\cite{Cook:2007} as a symmetry-based alternative useful when a reasonably well-defined approximate rotational symmetry is still present. The AKV method works directly on the structured simulation grid $(\theta_{\rm sim}, \phi_{\rm sim})$, without the intermediate triangulation and Ricci flow used in CMM. It identifies a preferred axial direction by seeking a divergence-free vector field $\phi^A$ on $\mathcal{S}$ that most nearly satisfies Killing's equation, minimizing the $L^2$-norm of the symmetric gradient $S_{AB} = D_{(A}\phi_{B)}$ subject to a normalization constraint. This leads to a generalized eigenvalue problem whose smallest eigenvalue measures the degree of axisymmetry violation. The associated stream function foliates the surface into flow lines, from which an invariant latitude $\theta_{\rm AKV}$ is constructed via cumulative area fractions; an associated azimuthal coordinate $\phi_{\rm AKV}$ is obtained from the AKV-flow transit time along each flow line. Because the resulting coordinate prescription $(\theta_{\rm AKV}, \phi_{\rm AKV})$ is conceptually different from $(\theta_{\rm CMM}, \phi_{\rm CMM})$, the AKV method yields, in general, a different set of multipole moments.

Both implementations are validated against the Kerr benchmarks summarized in Sec.~\ref{sec:theory}; the comparison is presented in Sec.~\ref{sec:kerr}.

\section{Validation against analytic Kerr solutions}
\label{sec:kerr}

To validate the numerical pipeline described above, we apply it to analytic Kerr black-hole horizons, using the Kerr formulation of the conformal framework developed in Sec.~\ref{sec:theory} as our benchmark. The Kerr horizon provides an ideal test case: its metric is known analytically, yet for nonzero spin the conformal colatitude~$\theta_{\rm CMM}$, defined by $Z=\cos\theta_{\rm CMM}$ in the notation of Sec.~\ref{sec:theory}, differs nontrivially from the Boyer--Lindquist colatitude~$\theta_{\rm BL}$. Thus the test checks not only the final multipole integrals, but also the reconstruction of the conformal map itself, exercising Ricci flow, spectral embedding, and gauge fixing. We compute the multipole moments for spin parameters $a/M \in [0,1]$ and compare the numerical results with the Kerr benchmarks summarized in Sec.~\ref{sec:theory}.

\begin{figure}
\includegraphics[width=\columnwidth]{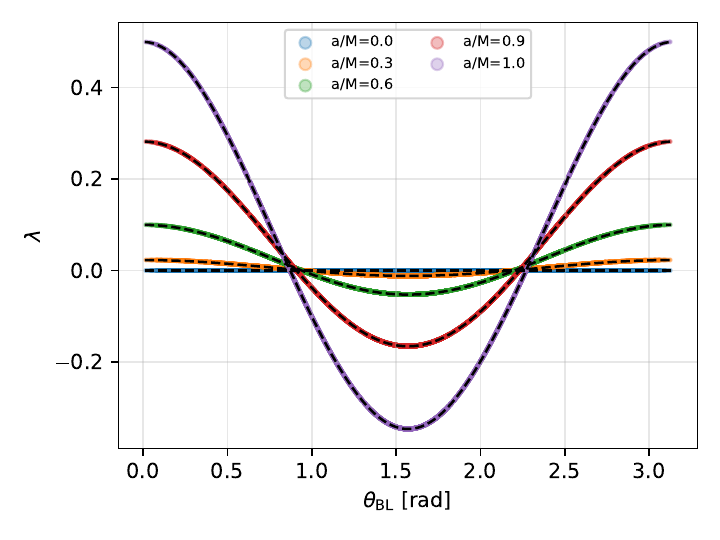}
\caption{\label{fig:conformal_factor} Logarithmic conformal factor $\lambda=\ln(R_{\Delta}\psi)$ of the black-hole horizon as a function of the Boyer--Lindquist colatitude for various spin parameters $a/M$. Colored dots represent CMM results, while dashed lines show the analytic Kerr values.}
\end{figure}

\begin{figure*}
    \includegraphics[width=0.55\textwidth]{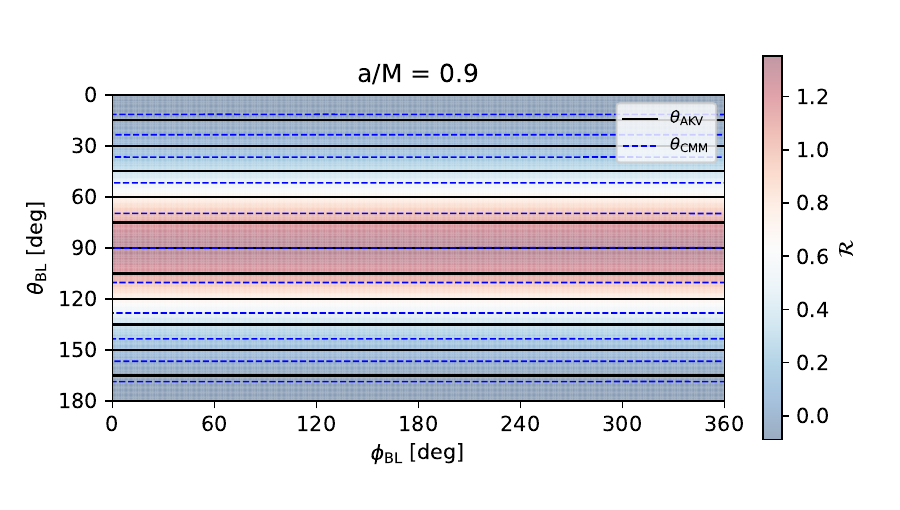}
    \hspace{-0.5cm}
    \includegraphics[width=0.42\textwidth]{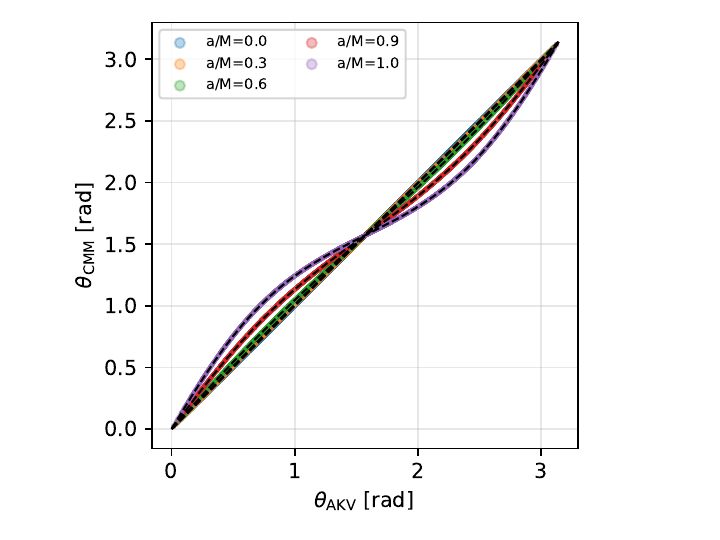}
    \caption{\label{fig:colatitude_mapping} Comparison of the colatitudes determined by the AKV method and CMM on Kerr horizons. \emph{Left:} contours of the AKV colatitude $\theta_{\rm AKV}$ (solid black) and the conformal colatitude $\theta_{\rm CMM}$ (dashed blue), spaced at $15^{\circ}$ intervals, on the Boyer--Lindquist parameter plane $(\phi_{\rm BL}, \theta_{\rm BL})$ for $a/M=0.9$; the background color encodes the intrinsic Ricci scalar $\mathcal{R}$ of the horizon. Since the Kerr horizon is axisymmetric, the AKV colatitude coincides with the Boyer--Lindquist one, $\theta_{\rm AKV}=\theta_{\rm BL}$. \emph{Right:} the conformal colatitude $\theta_{\rm CMM}$ as a function of $\theta_{\rm AKV}$ for $a/M\in\{0,\,0.3,\,0.6,\,0.9,\,1.0\}$. Colored dots show CMM results, whereas the dashed black curves correspond to the analytic Kerr maps.}
\end{figure*}

Fig.~\ref{fig:conformal_factor} shows the logarithmic conformal factor $\lambda$ as a function of the Boyer--Lindquist colatitude $\theta_{\rm BL}$ for Kerr horizons with spins $a/M\in\{0, 0.3, 0.6, 0.9, 1.0\}$. This is the numerical counterpart of Fig.~9 of Ref.~\cite{Gourgoulhon:2026}, where the same quantity is plotted analytically and denoted $E=\ln(R_{\Delta}\psi)$. The CMM results (colored dots) agree with the analytic Kerr benchmarks (dashed lines) across the full range of spins. The logarithmic conformal factor $\lambda$ vanishes identically for Schwarzschild ($a/M=0$) and grows in amplitude with increasing spin, reflecting the increasing deviation of the intrinsic horizon geometry from that of a round sphere.

Fig.~\ref{fig:colatitude_mapping} compares the colatitudes determined by the AKV and conformal constructions on Kerr horizons. The left panel overlays level sets of $\theta_{\rm AKV}$ (solid black) and $\theta_{\rm CMM}$ (dashed blue) on the Boyer--Lindquist coordinate plane $(\phi_{\rm BL}, \theta_{\rm BL})$ for $a/M = 0.9$, with the intrinsic Ricci scalar $\mathcal{R}$ shown as the background color. For the axisymmetric Kerr geometry, the AKV construction recovers the Boyer--Lindquist colatitude, $\theta_{\rm AKV}=\theta_{\rm BL}$, as expected. The conformal colatitude, however, gives a different parametrization of the same axial orbit space, as discussed in Sec.~\ref{sec:theory}. This difference is visible in the left panel: the conformal contours agree with the AKV contours at the poles and at the equator, but are displaced at intermediate latitudes, where the remapping between $\theta_{\rm BL}$ and $\theta_{\rm CMM}$ is largest. The same pattern is shown in the right panel, which plots $\theta_{\rm CMM}$ as a function of $\theta_{\rm AKV}$ for $a/M \in \{0, 0.3, 0.6, 0.9, 1.0\}$: the curves pass through $(0,0)$, $(\pi/2, \pi/2)$, and $(\pi, \pi)$, with deviations from the diagonal that grow with spin. The numerical results (colored dots) agree with the analytic Kerr maps (dashed black) across all five spins. 

\begin{figure}    \includegraphics[width=\columnwidth]{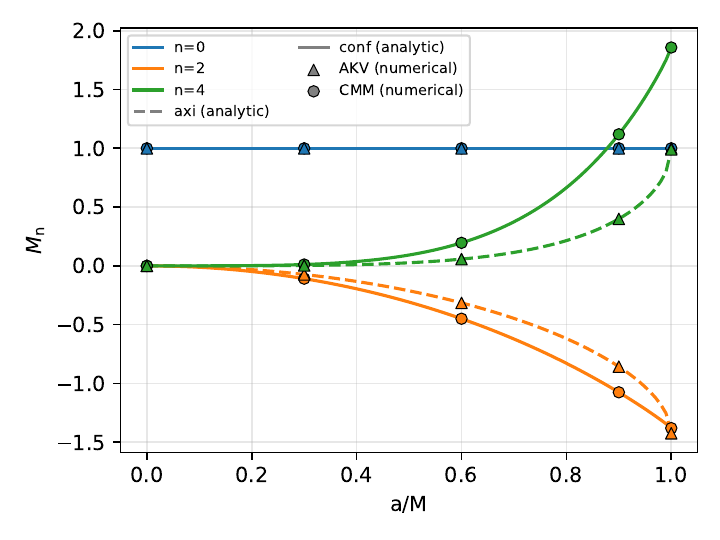}    \caption{\label{fig:Mn_vs_spin} Mass multipole moments $M_{n}$ of the Kerr horizon as functions of $a/M$ for even $n=0,\,2,\,4$. Solid and dashed curves show the analytic Kerr benchmarks for the conformal and axisymmetric definitions, respectively; circles and triangles show the results from CMM and the AKV method, respectively.} 
\end{figure}

Fig.~\ref{fig:Mn_vs_spin} compares the mass multipole moments $M_n$ obtained from our numerical implementations of the AKV method and CMM (triangles and circles, respectively) with the analytic Kerr values (dashed curves for the axisymmetric construction, solid curves for the conformal construction), plotted as functions of $a/M$ for even $n = 0, 2, 4$; the odd-$n$ moments vanish by equatorial symmetry and are therefore omitted from the figure. The analytic curves are obtained from the dimensionless multipoles defined in Sec.~\ref{sec:theory} [Eqs.~\eqref{eq:theory_axi_1d} and~\eqref{eq:theory_conf_axisym_1d}], converted to the dimensionful $M_n$ via Eqs.~\eqref{eq:theory_axi_Mn_Jn} and~\eqref{eq:theory_conf_Mlm_Jlm}; the corresponding dimensionless plot can be found in Fig.~15 (left panel) of Ref.~\cite{Gourgoulhon:2026}. At $n = 0$, both methods give $M_0 = M$, the horizon mass, which is set to unity here. For $n \geq 2$, the two methods give different values; the conformal moments are larger in magnitude over most of the spin range shown, except for the $n=2$ mode near $a/M=1$. The AKV triangles and CMM circles fall on their respective analytic benchmarks, confirming the expected relation between the axisymmetric and conformal Kerr multipoles. Together with Figs.~\ref{fig:conformal_factor} and~\ref{fig:colatitude_mapping}, this demonstrates that CMM correctly reconstructs both the Kerr conformal map and the associated multipoles.

\section{Application to binary black hole simulations}
\label{sec:bbh}

Having validated CMM against the analytic Kerr benchmarks in the preceding section, we now apply it, together with the AKV method, to a dynamical binary-black-hole spacetime. As a first dynamical application of the framework, we consider the simplest case: an equal-mass, non-spinning binary on quasi-circular orbits.

\begin{figure*}
  \centering
  \includegraphics[width=\linewidth]%
    {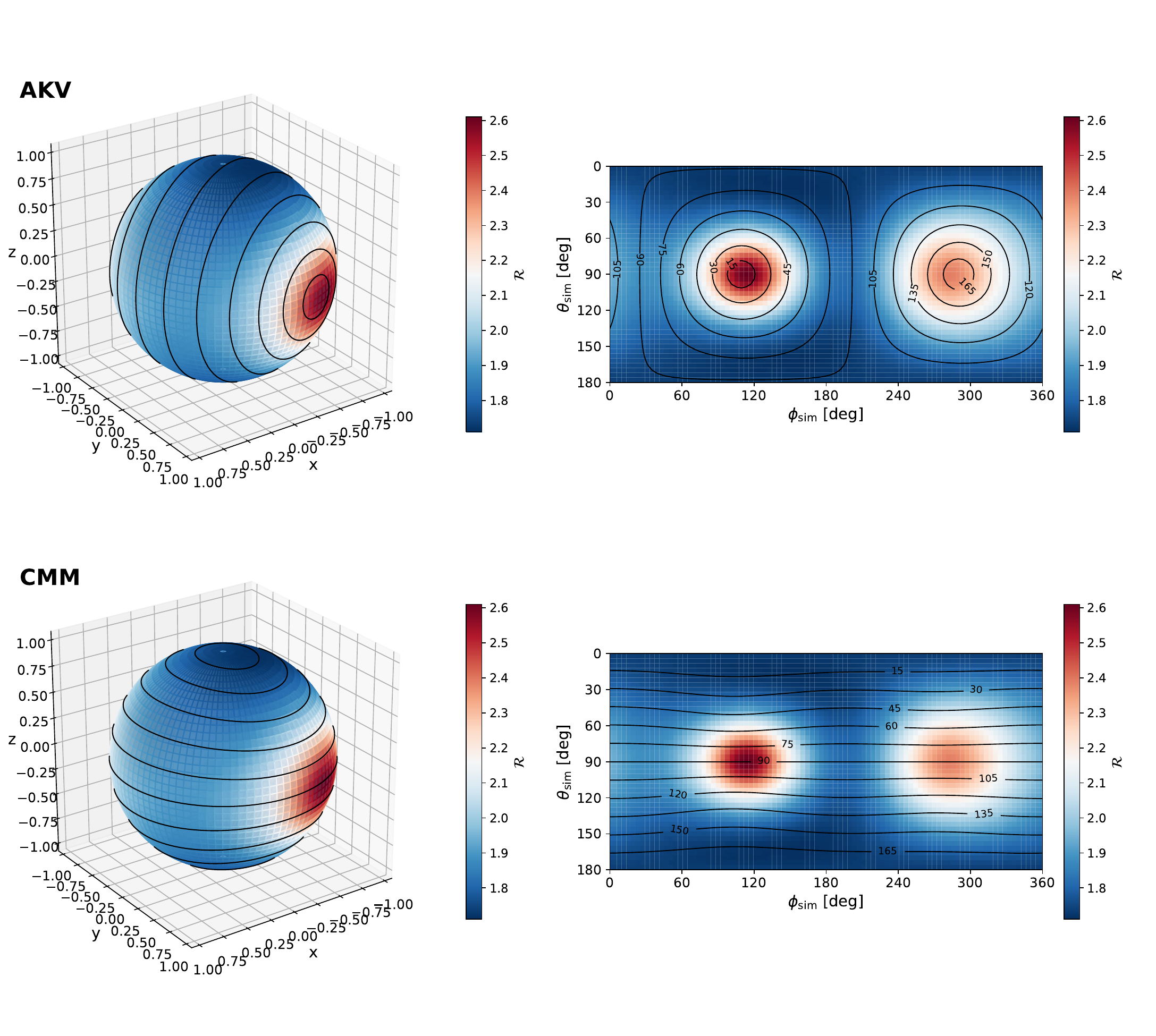}
  \caption{%
    Colatitude contours from the AKV method (top row) and CMM
(bottom row) on an individual binary-black-hole horizon at
    simulation time $t/M = 1328$, just before merger.  The left column shows the horizon as a unit sphere parametrized by the simulation grid
    $(\theta_{\rm sim},\,\phi_{\rm sim})$ and colored by the intrinsic
    Ricci scalar $\mathcal{R}$, with black level sets of
    $\theta_{\rm AKV}$ (top) or $\theta_{\rm CMM}$ (bottom) overlaid;
    the right column shows the same Ricci map and contours on the
    simulation coordinate plane $(\phi_{\rm sim},\,\theta_{\rm sim})$.%
  }
  \label{fig:ricci-theta-contours}
\end{figure*}

The simulation is performed with the Einstein Toolkit~\cite{EinsteinToolkit:2025,Loffler:2012}, using the TwoPunctures thorn~\cite{Ansorg:2004} for Bowen--York puncture initial data and the McLachlan thorn~\cite{Brown:2009} for BSSN evolution with $1+\log$ slicing and $\Gamma$-driver shift conditions. The computational domain employs the Carpet adaptive mesh refinement (AMR) driver~\cite{Schnetter:2004} with the Llama multipatch infrastructure~\cite{Pollney:2011}. The initial configuration follows the equal-mass model of Ref.~\cite{Prasad:2022}, with the finest AMR grid spacing of $M/64$.

At each time step, apparent horizons are located using AHFinderDirect~\cite{Thornburg:2004}, and quasilocal quantities on the horizons are computed by the QuasiLocalMeasures thorn~\cite{Dreyer:2003}. The horizon angular grid has a resolution of $(n_\theta, n_\phi) = (49, 96)$, including ghost zones. During the inspiral, individual horizons are tracked as surfaces~1 and~2, denoted BH1 and BH2, while the post-merger common horizon is tracked as surface~3.  Multipole moments are computed in post-processing from the VTK surface data using the AKV method and the CMM described in Sec.~\ref{sec:implementation}.

\begin{figure*}
  \centering
  \includegraphics[width=\linewidth]{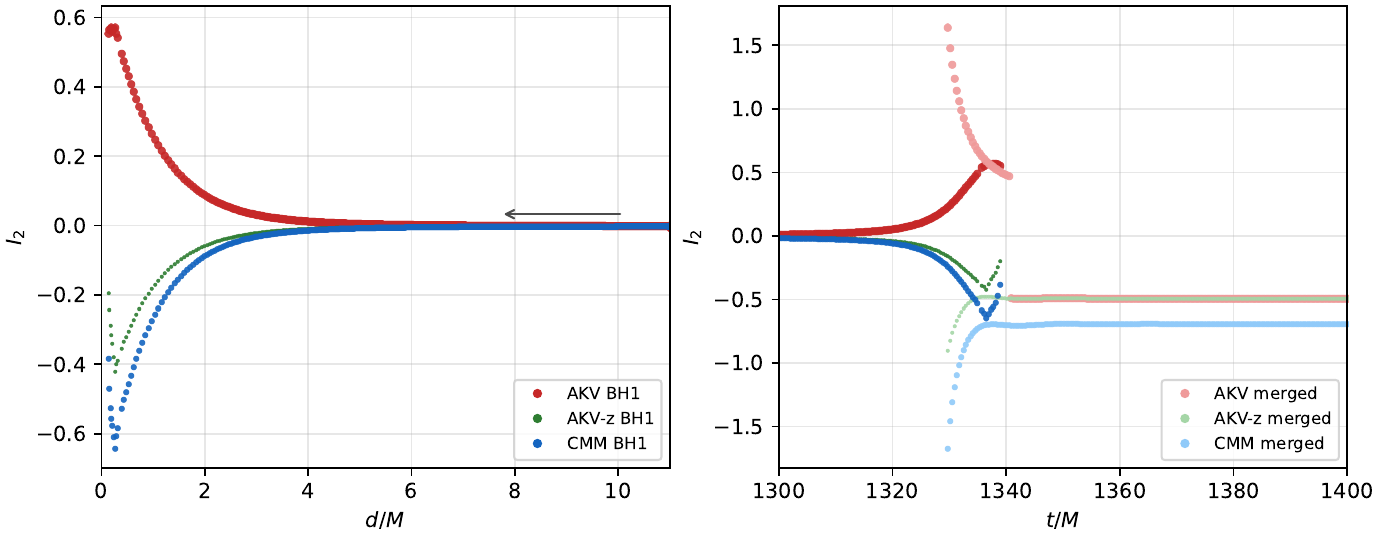}
  \caption{%
    Dimensionless mass quadrupole moment $I_2$ computed with the AKV method, the AKV-$z$ diagnostic, and the CMM. Here AKV-$z$ denotes a rotation of the AKV result into a frame whose approximate symmetry axis is aligned with the simulation $z$ axis. \emph{Left:} $I_2$ for one of the equal-mass black holes (BH1) as a function of the coordinate separation between the two black holes, $d/M$. \emph{Right:} Time evolution of $I_2$ for BH1 and for the remnant black hole. In this figure, red, green, and blue points denote AKV, AKV-z, and CMM, respectively. Dark and light shades distinguish BH1 before merger and the merged remnant, respectively.
  }
  \label{fig:I2_distance_time}
\end{figure*}

In contrast to the isolated Kerr horizons of Sec.~\ref{sec:kerr}, the individual horizons in a binary black-hole spacetime are subject to the gravitational influence of the companion. Each individual horizon is tidally elongated toward the other, giving it an approximate axis of symmetry along the companion direction. The AKV method correspondingly identifies a symmetry axis close to this direction, as illustrated by the $\theta_{\rm AKV}$ contours overlaid on the intrinsic Ricci scalar $\mathcal{R}$ in the top row of Fig.~\ref{fig:ricci-theta-contours}. The time-dependent tidal field associated with the orbital motion, however, breaks the exact axisymmetry of the horizon, with the departure growing as the binary tightens. The AKV axis is therefore offset from the instantaneous companion direction, and the AKV $\theta$-contours are slightly misaligned with the level sets of $\mathcal{R}$.

Across merger, the AKV-extracted axis changes abruptly: during the inspiral it tracks the companion direction and rotates with the binary, while after common-horizon formation the remnant settles toward a Kerr-like state whose symmetry axis is aligned with the orbital angular momentum direction---the simulation $z$ axis for our non-spinning, equal-mass binary. The AKV axis therefore changes from an in-plane direction to the $z$ axis, and the AKV multipole moments inherit this discontinuity. For an illustrative comparison in the simulation $z$-aligned frame, we also consider a diagnostic set of multipoles, denoted AKV-$z$, obtained by applying a Wigner-\(D\) rotation to the AKV-frame multipole coefficients using the orientation of the AKV axis in the simulation coordinates,
\begin{equation}
I^{\mathrm{AKV}\text{-}z}_{\ell m} = \sum_{m'} D^{\ell}_{m' m}(\alpha, \beta, \gamma)\, I^{\mathrm{AKV}}_{\ell m'},
\end{equation}
where the Euler angles $(\alpha, \beta, \gamma)$ are chosen so that the AKV symmetry axis is rotated onto the simulation $z$ axis. This AKV-$z$ construction is used as a diagnostic comparison; the residual ambiguity in rotating an AKV frame into the simulation frame is discussed further in Sec.~\ref{sec:discussion}.

The CMM, in contrast, does not extract an axis from the horizon geometry: it yields a round-sphere metric defined only up to a global \(\mathrm{SO}(3)\)  rotation, and the orientation must be specified by an external convention. We adopt the simulation $z$ axis as the polar axis of the round-sphere coordinates throughout the simulation, and fix the azimuthal origin by mapping the simulation $x$ axis onto the $\varphi_{\rm conf} = 0$ meridian, as described in Sec.~\ref{sec:implementation}. With this convention, the reference frame does not undergo the abrupt reorientation seen in the AKV construction. The bottom row of Fig.~\ref{fig:ricci-theta-contours} shows the corresponding conformal colatitude contours; their spacing becomes denser in regions of larger $\mathcal{R}$, reflecting the tendency of the conformal map to redistribute the round-sphere coordinates toward regions of higher intrinsic curvature.

Fig.~\ref{fig:I2_distance_time} shows the dimensionless quadrupole moment $I_2$ computed using the AKV method, the AKV-$z$ diagnostic, and CMM, as functions of the coordinate separation $d/M$ and the simulation time $t/M$. Here $d$ is the Euclidean distance between the horizon centroids in the simulation frame. Although this is not a gauge-invariant proper distance, it has been shown to agree well with estimates based on the instantaneous gravitational-wave frequency or post-Newtonian orbital separation~\cite{Prasad:2022}. The coordinate separation should therefore be interpreted as a useful separation parameter for tracking the dynamical tidal response, although its meaning becomes less clear near common-horizon formation.

The sign difference between the AKV result and the $z$-aligned AKV-$z$ and CMM results reflects the different frames in which the quadrupolar deformation is decomposed. During the inspiral, the tidal field of the companion stretches each individual horizon approximately along the direction toward the other black hole. Since the AKV coordinate system is adapted to this companion-facing direction, the deformation appears prolate in the AKV frame, giving a positive value of $I_{2}$. By contrast, in the AKV-$z$ and CMM frames the polar axis is aligned with the orbital angular-momentum direction. In these frames the tidal bulge lies mainly in the equatorial plane, so the same deformation has an oblate-like appearance and $I_2$ becomes negative. Although the individual horizons are not exactly axisymmetric during inspiral, this frame-dependent prolate/oblate picture explains the opposite signs of $I_2$ in the different frames.

\begin{figure*}
    \includegraphics[width=\textwidth]{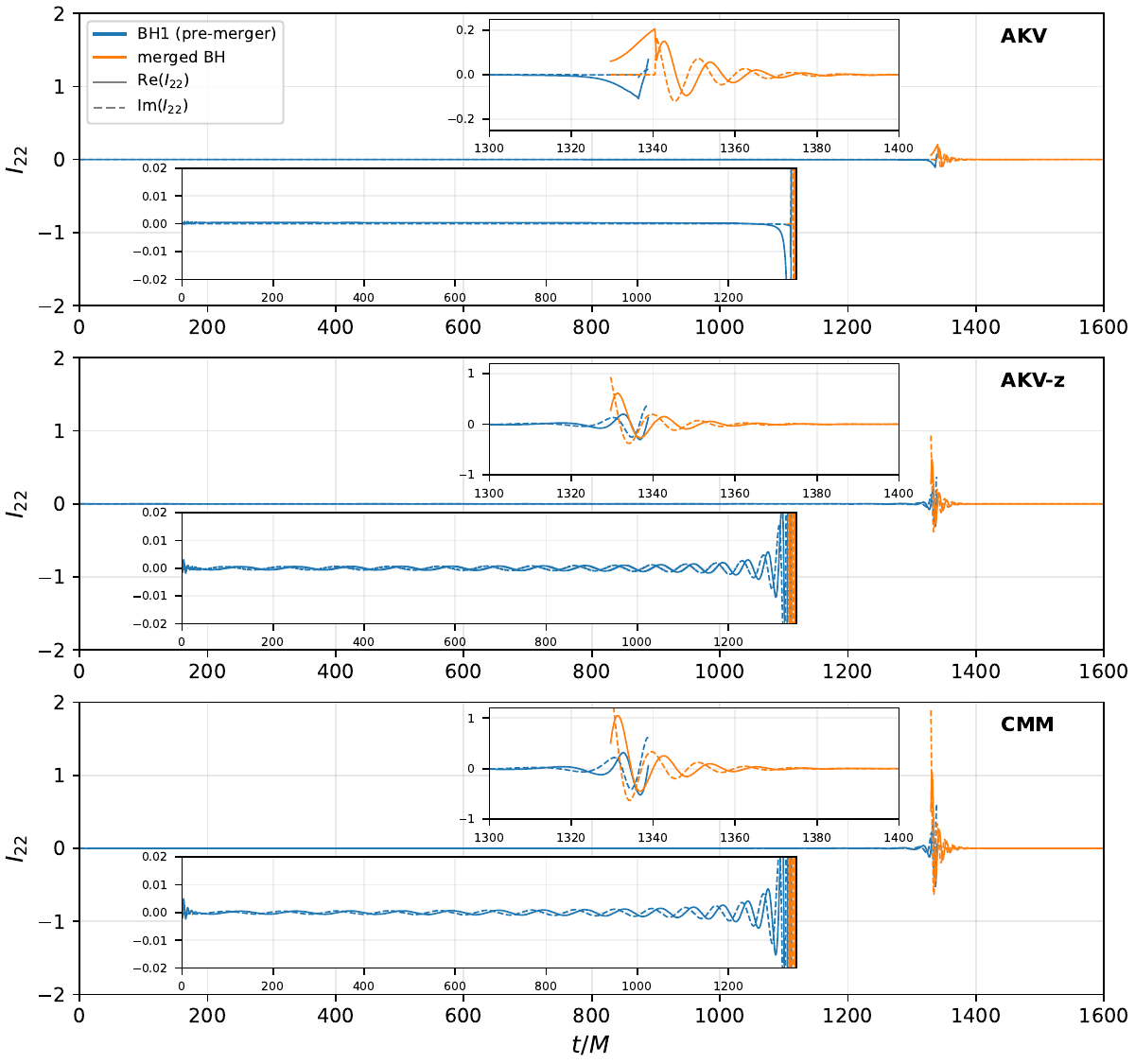}
    \caption{\label{fig:I22_comparison} Time evolution of the geometric multipole moment $I_{22}$ for an equal-mass, non-spinning binary black hole merger, computed with the AKV method (top), the AKV-$z$ diagnostic (middle), and the CMM (bottom). Blue curves correspond to the individual horizon BH1 before merger, and orange curves to the common horizon after merger. Solid and dashed lines represent the real and imaginary parts of $I_{22}$, respectively. Insets provide zoomed views of the indicated time intervals.}
\end{figure*}

The right panel of Fig.~\ref{fig:I2_distance_time} shows the subsequent evolution through merger and ringdown. The values shown for BH1 and the common horizon are evaluated on distinct surfaces and are therefore not required to agree at common-horizon formation. The AKV result is additionally affected by the abrupt reorientation of the approximate symmetry axis near common-horizon formation. During inspiral, this axis is approximately tied to the companion-facing direction, whereas after merger it aligns with the orbital angular momentum as the common horizon relaxes toward a Kerr-like remnant. Consequently, after this transition the AKV and AKV-$z$ results coincide. Before reaching their final plateau values, the multipoles exhibit a short period of damped oscillatory behavior, reflecting the ringing of the newly formed common horizon. At late times, both the AKV and CMM results settle to approximately constant negative values, but the asymptotic values are different because the two methods use different coordinate systems, and hence different spherical-harmonic bases, even for an axisymmetric Kerr horizon. For an equal-mass, non-spinning binary, the final dimensionless spin of the remnant is known to be $a/M \simeq 0.686$ \cite{Scheel:2009}. Substituting this value into the analytic Kerr benchmarks gives $I_{2}^{\textrm{axi}}\simeq-0.489$ for the axisymmetric/AKV definition and $I_{2}^{\textrm{conf}}\simeq-0.684$ for the conformal definition. These values are consistent with the late-time plateaus of the AKV and CMM curves, respectively, indicating that both methods approach the appropriate Kerr limits during ringdown. 

Fig.~\ref{fig:I22_comparison} shows the time evolution of the complex quadrupole mode $I_{22}$ for the same equal-mass, non-spinning binary. The qualitative behavior of this mode depends strongly on the frame used in the multipole decomposition. In the AKV frame, the polar axis during inspiral is approximately tied to the companion-facing direction. Since the dominant tidal deformation is mostly axisymmetric about this direction, it contributes mainly to the $m=0$ quadrupole component, leaving $I_{22}$ close to zero during most of the inspiral. Toward the end of the inspiral, however, $I_{22}$ starts to deviate from zero as the horizon geometry becomes increasingly non-axisymmetric near merger.

In contrast, the AKV-$z$ and CMM frames are aligned with the orbital angular momentum. In these frames the companion-induced tidal bulge lies primarily in the orbital plane and appears as a non-axisymmetric $m=2$ deformation. The corresponding $I_{22}$ amplitudes grow during inspiral, and their instantaneous frequencies increase as the binary tightens. After common-horizon formation, the amplitudes decay as the remnant rings down toward an axisymmetric Kerr horizon. Thus, in the $z$-aligned frames, the horizon quadrupole mode exhibits a qualitative inspiral--merger--ringdown pattern similar to that of the gravitational waveform.

Near merger, the pre-merger signal from the individual horizon and the post-merger signal from the common horizon overlap during a short time interval and show a relatively smooth transition. This suggests that the quadrupolar distortion of the individual horizons is, in a qualitative sense, carried into the newly formed common horizon. The conformal frame is especially well suited for this comparison because its orientation is fixed externally and maintained consistently across merger. A direct comparison between these horizon multipoles and the gravitational waveform is left for future work.

\section{Summary and Discussion}
\label{sec:discussion}

In this work, we have presented the conformal-mapping method (CMM), the first numerical implementation of the conformal
construction for horizon multipoles without assuming axisymmetry. The method constructs a canonical unit round metric in the conformal class of the physical horizon metric, fixes the M\"obius freedom by imposing the vanishing-area-dipole condition, and then computes multipole moments in the associated spherical-harmonic basis. Numerically, this was realized through a pipeline based on discrete Ricci flow, spectral embedding, and gauge fixing of the residual conformal freedom. We validated the implementation against analytic Kerr benchmarks and applied it to an equal-mass, non-spinning binary-black-hole merger, comparing the results with those obtained from the AKV method.

The binary application highlights the central distinction between symmetry-based and conformal approaches. A symmetry-based method, represented here by the AKV construction, adapts its coordinates to an approximate symmetry axis of the horizon, while CMM allows the multipoles to be expressed in an externally fixed frame, such as one aligned with the orbital angular momentum. As a result, the two types of frame choices emphasize different aspects of the same horizon deformation. In the symmetry-adapted frame, the tidal distortion of an individual horizon is largely captured as an approximately axisymmetric quadrupole about the companion-facing direction. In the $z$-aligned AKV-$z$ and CMM frames, the same distortion appears instead as a non-axisymmetric $m=2$ deformation in the orbital plane. This frame dependence is essential for interpreting the signs and amplitudes of the multipole moments during the inspiral--merger--ringdown evolution.

The main advantage of such symmetry-based prescriptions is that they identify an approximate symmetry axis directly from the horizon geometry. When the horizon is close to axisymmetric, this provides a natural coordinate system adapted to the intrinsic geometry of the surface. This is particularly useful for isolated or nearly isolated spinning black holes, late-time remnant horizons, or configurations in which the horizon distortion has a single dominant direction. In such cases, the symmetry-adapted frame gives a clear geometric interpretation of the multipoles: the $m=0$ modes describe the axisymmetric part of the deformation about the selected axis, while the $m\neq0$ modes measure deviations from that approximate symmetry.

This same symmetry-adapted character, however, becomes a limitation in dynamical situations. Since the symmetry-adapted frame follows the instantaneous approximate symmetry axis, the coordinate system can change in time as the dominant deformation changes. In the binary simulation studied here, the symmetry axis of an individual horizon is approximately tied to the companion-facing direction during inspiral, but after common-horizon formation it aligns with the orbital angular momentum direction as the remnant relaxes toward a Kerr-like state. This change of axis produces the jump seen in the AKV multipoles near merger. It also makes the AKV modes less convenient for tracking a time-dependent horizon signal in a fixed physical frame, such as the frame normally used to decompose gravitational waveforms.

The physical meaning of a single AKV axis can also become less clear when several sources of distortion are present. For example, spin-induced distortions are naturally associated with the spin direction, whereas tidal distortions are associated with the companion direction. If these directions are misaligned, the horizon need not possess a single physically preferred axis, even approximately. The symmetry-based method will still select the vector field that best approximates a rotational symmetry, but the resulting axis may mix distinct physical effects and may not by itself provide a simple interpretation of the multipole content.

The AKV-$z$ multipoles used in Sec.~\ref{sec:bbh} should therefore be interpreted as a useful diagnostic rather than as a fully specified coordinate prescription. The AKV method determines an approximate symmetry axis, but it does not, by itself, provide a complete map between the directions in the simulation frame and those in the AKV coordinate system. In particular, in a generic non-axisymmetric situation it is not obvious where the simulation $x$ and $z$ directions should be placed in the AKV coordinates. This makes rotations of AKV multipoles into externally specified simulation frames less straightforward. Nevertheless, the AKV-$z$ diagnostic is useful in the present equal-mass, non-spinning case because it illustrates how the same AKV multipole content changes when viewed in a frame aligned with the orbital angular momentum.

The CMM takes a different approach. It does not attempt to identify a symmetry axis from the horizon geometry. Instead, once the canonical conformal round metric has been fixed, the remaining freedom is only a global \(\mathrm{SO}(3)\)  rotation of the round sphere. This residual orientation can be fixed by an external convention. In this work, we chose the simulation $z$-axis, aligned with the orbital angular momentum, as the polar axis and used a reference simulation direction to fix the azimuthal origin. With this choice, the conformal multipoles are expressed in a frame that remains fixed throughout inspiral, merger, and ringdown.

This fixed-frame property is particularly useful for studying dynamical horizon multipoles. In the conformal frame, the $I_{22}$ mode shows the qualitative inspiral--merger--ringdown behavior expected for a quadrupolar dynamical signal: its amplitude and instantaneous frequency grow during inspiral, and the amplitude decays after common-horizon formation as the remnant rings down toward a Kerr horizon. Because gravitational-wave modes are also decomposed in an externally specified asymptotic frame, the conformal multipoles provide a natural framework for future comparisons between horizon dynamics and gravitational radiation, with the remaining rotational convention fixed consistently between the horizon and waveform frames.

This residual \(\mathrm{SO}(3)\)  freedom is not only a limitation but also a source of useful flexibility, provided that the orientation convention is specified consistently: different physical questions may call for different frames, such as a spin-aligned frame for the remnant or a companion-aligned frame for the tidal deformation of an individual horizon. Once the conformal multipoles have been computed in one chosen frame on the conformal unit round sphere, they can be re-expressed in any other frame obtained by an \(\mathrm{SO}(3)\)  rotation on the same round sphere using the usual Wigner-\(D\) transformation within each $\ell$ sector. In this sense, the conformal construction fixes the canonical round metric and the associated harmonic decomposition up to an overall rotation, while leaving the final orientation on the round sphere to be chosen by an external convention.


Moreover, CMM does not require a preferred axis as an input, but preferred directions can still be identified a posteriori from the multipoles themselves. A natural starting point is the \(\ell=2\) mass sector, which may be assembled into a symmetric trace-free (STF) tensor
\begin{equation}
Q_{ij}
 :=
 \sum_{m=-2}^{2}
 I^{(\mathrm{conf})}_{2m}\,
 \mathcal{Y}^{2m}_{ij},
\end{equation}
where \(\mathcal{Y}^{2m}_{ij}\) denotes an STF tensor basis on the conformal round sphere. The principal directions on this round sphere are then obtained from the eigenvalue problem of \(Q^{i}{}_{j}\). If the quadrupolar sector is exactly axisymmetric, then the nondegenerate eigenvector of \(Q_{ij}\) determines its symmetry axis up to sign. In the rotated frame aligned with that direction, the quadrupole takes the standard axis-adapted form, with only the \(m=0\) component remaining and all \(m\neq0\) components vanishing. More generally, the eigensystem of \(Q_{ij}\) provides the principal directions of the net quadrupolar deformation and may help interpret situations in which spin-induced and tidally induced distortions are not aligned.

The waveform-like behavior of the $z$-aligned horizon multipoles suggests a natural direction for future work. In the present paper, we have focused on the formulation, implementation, and basic numerical tests of the CMM, and have not attempted a direct quantitative comparison with the gravitational radiation. A detailed comparison of amplitudes, phases, and instantaneous frequencies between the horizon multipoles and the gravitational waveform, with the frame conventions treated consistently, is left for future work. It would also be useful to apply the method to more general binary configurations, such as unequal-mass and spinning binaries, where the individual horizons experience different tidal and spin-induced distortions. Close encounters and scattering configurations would provide another setting in which to test the robustness of the conformal construction under strong, transient horizon distortions.


\begin{acknowledgments}
This work was supported by the National Research Foundation of Korea (NRF) grants funded by the Korea government (MSIT) (RS-2025-00556091 and RS-2025-00564350). Computing resources and technical support are provided by the IBS Research Solution Center, KISTI Supercomputing Center (No. KSC-2024-CRE-0374), and gmunu HPC cluster of KASI. We thank APCTP, Pohang, Korea for their hospitality during the Program Type APCTP-2026-T01 and APCTP-2026-S09 from which this work greatly benefited.  
\end{acknowledgments}

\bibliography{ref}

\end{document}